\documentclass{article}
\usepackage{amssymb}
\usepackage{amsmath}
\usepackage{hyperref}

\input{tcilatex}
\begin{document}

\title{Massive Dual Spinless Fields Revisited}
\author{Thomas L. Curtright \\
Department of Physics, University of Miami\\
P O Box 248046, Coral Gables, Florida 33124}
\date{21 August 2019}
\maketitle

\begin{abstract}
Massive dual spin zero fields are reconsidered in four spacetime dimensions.
\ A closed-form Lagrangian is presented that describes a field coupled to
the curl of its own energy-momentum tensor.

\noindent \hrulefill
\end{abstract}

\bigskip

\begin{center}
In tribute to \href{https://en.wikipedia.org/wiki/Peter_Freund}{Peter George
Oliver Freund} (1936-2018)
\end{center}

\bigskip

\section*{Introduction}

As indicated in the Abstract, the point of this paper is to find an explicit
Lagrangian for the dual form of a massive scalar field self-coupled in a
particular way to its own energy-momentum tensor. \ This boils down to a
well-defined mathematical problem whose solution is given here, thereby
completing some research initiated and published long ago in this journal 
\cite{CF1980}.

After first presenting a concise mathematical statement of the problem, and
then giving a closed-form solution in terms of elementary functions, the
field theory that led to the problem is re-examined from a fresh
perspective. \ The net result is a very direct approach that leads to both
the problem and its solution. \ 

\section*{Some History}

Here I reconsider research first pursued in collaboration with Peter Freund,
in an effort to tie up some loose ends. \ In the spring of 1980, when I was
a post-doctoral fellow in Yoichiro Nambu's theory group at The Enrico Fermi
Institute, Peter and I were confronted by a pair of partial differential
equations (see \cite{CF1980} p 417).%
\begin{eqnarray}
m^{2}\frac{\partial ^{2}\mathcal{L}}{\partial v^{2}}+2\left( \frac{\partial 
\mathcal{L}}{\partial u}\right) &=&4g\left( \frac{\partial \mathcal{L}}{%
\partial u}\right) \left( u\frac{\partial ^{2}\mathcal{L}}{\partial
u\partial v}+v\frac{\partial ^{2}\mathcal{L}}{\partial v^{2}}-\frac{\partial 
\mathcal{L}}{\partial v}\right) \ ,  \label{1st} \\
m^{2}\frac{\partial ^{2}\mathcal{L}}{\partial u\partial v} &=&4g\left( \frac{%
\partial \mathcal{L}}{\partial u}\right) \left( u\frac{\partial ^{2}\mathcal{%
L}}{\partial u^{2}}+v\frac{\partial ^{2}\mathcal{L}}{\partial u\partial v}-%
\frac{\partial \mathcal{L}}{\partial u}\right) \ ,  \label{2nd}
\end{eqnarray}%
where $m$ and $g$ are constants. \ We noticed in passing that these PDEs
imply the secondary condition \cite{WH}%
\begin{equation}
\left( \frac{\partial ^{2}\mathcal{L}}{\partial u\partial v}\right)
^{2}=\left( \frac{\partial ^{2}\mathcal{L}}{\partial u^{2}}\right) \left( 
\frac{\partial ^{2}\mathcal{L}}{\partial v^{2}}\right) \ ,  \label{3rd}
\end{equation}%
and we then looked for a solution to (1-3) as a series in $g$ beginning with 
\begin{equation}
\mathcal{L}\left( u,v\right) =\frac{1}{2}~v^{2}-\frac{1}{2}~m^{2}u+\frac{g}{m%
}\left( \frac{1}{3}~v^{3}-m^{2}uv\right) +O\left( g^{2}\right) \ .
\label{IC}
\end{equation}%
To simplify the equations\ to follow, I will rescale $g=m\kappa $ so that
the constant $m$ always appears in (1-3) only in the combination $v/m$. Thus
I may as well set $m=1$, and hence $\kappa =g$. \ I can then restore the
parameter $m$ in any subsequent solution for $\mathcal{L}$ by the
substitution $\mathcal{L}\left( u,v\right) \rightarrow m^{2}\mathcal{L}%
\left( u,v/m\right) $.

Clearly, there is a two-parameter family of exact solutions to these PDEs
which depends only on $v$, namely,%
\begin{equation}
\mathcal{L}_{0}\left( v\right) =a+b~v\ ,  \label{Topo}
\end{equation}%
where $a$ and $b$ are constants. \ However, for the model field theory that
gave rise to the partial differential equations (1,2), this linear function
of $v$\ amounts to a topological term in the action and therefore gives no
contribution to the bulk equations of motion. \ Moreover, $\mathcal{L}%
_{0}\left( v\right) $ contributes only a (cosmological) constant term to the
canonical energy-momentum tensor. \ So, in the context of our 1980 paper 
\cite{CF1980} where solutions of (1,2) were sought which gave more \emph{%
interesting} contributions, this $\mathcal{L}_{0}\left( v\right) $ was not
worth noting. \ Nevertheless, it reappeared in another context, somewhat
later \cite{FR}.

\section*{Completing Some Unfinished Business}

It so happened in 1980 that Peter and I did not find an exact $\mathcal{L}%
\left( u,v\right) $ to solve the PDEs (1,2). \ In fact, we reported then
only the terms given in (\ref{IC}). \ Here I wish to present an exact,
closed-form solution to all orders in $g$. \ 

The crucial feature leading to this particular solution is that the
dependence on $v$ is only through the linear combination $v-gu$. \ The
result is 
\begin{equation}
\mathcal{L}\left( u,v\right) =-\frac{1}{2}~u+\frac{1}{2}\left( v-gu\right)
^{2}+\frac{1}{3}~g\left( v-gu\right) ^{3}\left. _{3}F_{2}\right. \left( 1,%
\frac{1}{2},\frac{3}{2};2,\frac{5}{2};-4g^{2}\left( v-gu\right) ^{2}\right)
\ ,  \label{AnExactSolution}
\end{equation}%
where as a series%
\begin{gather}
F\left( w\right) \equiv \frac{1}{3}~w^{3}\left. _{3}F_{2}\right. \left( 1,%
\frac{1}{2},\frac{3}{2};2,\frac{5}{2};-4w^{2}\right) =\sum_{n=1}^{\infty }%
\frac{\left( 2n-2\right) !}{\left( n-1\right) !\left( n\right) !}\frac{%
\left( -1\right) ^{n+1}}{2n+1}~w^{2n+1}  \notag \\
=\frac{1}{3}~w^{3}-\frac{1}{5}~w^{5}+\frac{2}{7}~w^{7}-\frac{5}{9}~w^{9}+%
\frac{14}{11}~w^{11}-\frac{42}{13}~w^{13}+\frac{44}{5}~w^{15}+O\left(
w^{17}\right) \ .  \label{hyper}
\end{gather}%
Fortunately, the $\left. _{3}F_{2}\right. $ hypergeometric function in (\ref%
{hyper}) reduces to elementary functions. \ For real $w$, $\ $%
\begin{equation}
F\left( w\right) =-\frac{1}{2}~w+\frac{1}{4}~w\sqrt{1+4w^{2}}+\frac{1}{8}%
~\ln \left( 2w+\sqrt{1+4w^{2}}\right) \ .  \label{simple}
\end{equation}%
$\allowbreak $Nevertheless, the solution (\ref{AnExactSolution}) was first
obtained in its series form (\ref{hyper}) and only afterwards was it
expressed as a special case of the hypergeometric $\left. _{3}F_{2}\right. $%
, with its subsequent simplification to elementary functions.

More generally, it is not so difficult to establish that solutions to (1-3)
necessarily have the form%
\begin{equation}
\mathcal{L}\left( u,v\right) =-\frac{1}{2g}~v+G\left( v+2g\int^{u}H\left(
s\right) ds\right) \ ,  \label{GeneralForm}
\end{equation}%
where the function $G$ is differentiable, and $H$ is integrable, but
otherwise not yet determined, as befits the general solution of a more
easily solvable 1st-order PDE, albeit nonlinear:%
\begin{equation}
\frac{\partial }{\partial v}~\ln \left( \frac{\partial \mathcal{L}}{\partial
u}\right) =\frac{\partial }{\partial v}~\ln \left( 1+2g~\frac{\partial 
\mathcal{L}}{\partial v}\right) \ .  \label{Integrable2ndOrder}
\end{equation}%
Note in (\ref{GeneralForm}) the return of an explicit term linear in $v$. \
This term arises as the particular solution of the inhomogeneous 1st-order
PDE that results from integrating (\ref{Integrable2ndOrder}) and
exponentiating, namely,%
\begin{equation}
\frac{1}{H\left( u\right) }~\frac{\partial \mathcal{L}}{\partial u}-2g~\frac{%
\partial \mathcal{L}}{\partial v}=1\ .
\end{equation}%
The functions $G$ and $H$ are now constrained by additional conditions that
lie hidden within (1) and (2). \ 

I will leave it to the reader to flesh out those additional conditions. I
will not go through that analysis here. \ Instead, I will reconsider the
model field theory that led to the partial differential equations (1,2) in
light of the exact solution (\ref{AnExactSolution}). \ That solution
provides a good vantage point to view and analyze the model.

\section*{The Model Revisited}

Consider a Lagrangian density $\mathcal{L}\left( u,v\right) $ depending on a
vector field $V^{\mu }$ through the two scalar variables, 
\begin{equation}
u=V_{\mu }V^{\mu }\ ,\ \ \ v=\partial _{\mu }V^{\mu }\ .
\end{equation}%
This vector field is to be understood in terms of an antisymmetric, rank 3,
tensor gauge field, $V_{\alpha \beta \gamma }$, i.e. the four-dimensional
spacetime dual of a massive scalar \cite{CF1980}, with its corresponding
gauge invariant field strength, $F_{\mu \alpha \beta \gamma }=\partial _{\mu
}V_{\alpha \beta \gamma }-\partial _{\alpha }V_{\beta \gamma \mu }+\partial
_{\beta }V_{\gamma \mu \alpha }-\partial _{\gamma }V_{\mu \alpha \beta }$. \
Thus 
\begin{equation}
V^{\mu }=\frac{1}{6}~\varepsilon ^{\mu \alpha \beta \gamma }V_{\alpha \beta
\gamma }\ ,\ \ \ \partial _{\mu }V^{\mu }=\frac{1}{24}~\varepsilon ^{\mu
\alpha \beta \gamma }F_{\mu \alpha \beta \gamma }\ .  \label{VandF}
\end{equation}%
The bulk field equations that follow from the action of $\mathcal{L}\left(
u,v\right) $ by varying $V^{\mu }$ are simply%
\begin{equation}
\partial _{\mu }\mathcal{L}_{v}=2V_{\mu }\mathcal{L}_{u}\ ,
\label{FieldEqns}
\end{equation}%
where the partial derivatives of $\mathcal{L}$\ are designated by $\mathcal{L%
}_{u}\equiv \partial \mathcal{L}\left( u,v\right) /\partial u$ and $\mathcal{%
L}_{v}\equiv \partial \mathcal{L}\left( u,v\right) /\partial v$. \ An
obvious inference from these field equations is that the on-shell vector $%
V_{\mu }$ is a gradient of a scalar $\Phi $, 
\begin{equation}
V_{\mu }=\partial _{\mu }\Phi \ ,  \label{VectorAsGrad}
\end{equation}%
if and only if $\mathcal{L}_{u}$ is a function of $\mathcal{L}_{v}$. \ For
example, if $\mathcal{L}_{u}$ has a linear relation to $\mathcal{L}_{v}$
with $\mathcal{L}_{u}=a+b\mathcal{L}_{v}$ for constants $a$ and $b$, the
field equations give 
\begin{equation}
\Phi =\frac{1}{2b}\text{~}\ln \left( a+b\mathcal{L}_{v}\right) \ ,
\end{equation}%
More generally, if $\mathcal{L}_{u}=\Psi \left( \mathcal{L}_{v}\right) $,
then%
\begin{equation}
\Phi =\frac{1}{2}\int^{\mathcal{L}_{v}}\frac{dz}{\Psi \left( z\right) }\ .
\end{equation}%
But in any case, on-shell the combination $U_{\mu }=V_{\mu }\mathcal{L}_{u}$
is a spacetime gradient.

An additional gradient of the field equations then gives%
\begin{equation}
\partial _{\lambda }\partial _{\mu }\mathcal{L}_{v}=2\left( \partial
_{\lambda }V_{\mu }\right) \mathcal{L}_{u}+2V_{\mu }\partial _{\lambda }%
\mathcal{L}_{u}\ .
\end{equation}%
From $\partial _{\lambda }\partial _{\mu }\mathcal{L}_{v}=\partial _{\mu
}\partial _{\lambda }\mathcal{L}_{v}$ it follows that 
\begin{equation}
\left( \partial _{\mu }V_{\lambda }-\partial _{\lambda }V_{\mu }\right) 
\mathcal{L}_{u}=V_{\mu }\partial _{\lambda }\mathcal{L}_{u}-V_{\lambda
}\partial _{\mu }\mathcal{L}_{u}\ .
\end{equation}%
Thus the vector $V_{\mu }$ is a gradient of a scalar, as in (\ref%
{VectorAsGrad}), such that 
\begin{equation}
\partial _{\mu }V_{\lambda }=\partial _{\lambda }V_{\mu }\ ,
\label{GradientCond}
\end{equation}%
if and only if for some scalar function $\Omega $,%
\begin{equation}
\partial _{\lambda }\mathcal{L}_{u}=V_{\lambda }\Omega \ .  \label{Omega}
\end{equation}

\subsection*{Simplification}

Now for simplicity, demand that $\mathcal{L}_{u}=a+b\mathcal{L}_{v}$ for
constants $a$ and $b$, in accordance with $V_{\mu }$ being a gradient, as in
(\ref{VectorAsGrad}) and (\ref{GradientCond}). \ This linear condition is
immediately integrated to obtain%
\begin{equation}
\mathcal{L}\left( u,v\right) =au+L\left( v+bu\right) \ ,  \label{Lagrangian}
\end{equation}%
where $L\left( v+bu\right) $ is a differentiable function of the linear
combination $v+bu$. \ The field equations (\ref{FieldEqns}) are now%
\begin{equation}
\partial _{\lambda }\mathcal{L}_{v}=\partial _{\lambda }L^{\prime }=2\left(
a+bL^{\prime }\right) V_{\lambda }=2V_{\lambda }\mathcal{L}_{u}\ .
\label{SimpleFieldEqns}
\end{equation}%
That is to say, the scalar in (\ref{Omega}) is $\Omega =2ab+2b^{2}L^{\prime
}\ .$

\subsection*{Energy-momentum tensors}

In \cite{CF1980} Peter and I say that, given (1-3), the field equations for $%
V_{\mu }$ amount to (\ref{GradientCond}) along with the \textquotedblleft
simple, indeed elegant\textquotedblright\ statement%
\begin{equation}
\left( \square +m^{2}\right) V_{\mu }=\frac{g}{m}~\partial _{\mu }\theta \ ,
\label{Elegant}
\end{equation}%
where $g$ has units of length, and $\theta $ is the trace of the conformally
improved energy-momentum tensor. \ 

Be that as it may, there is a less oracular method to reach this form for
the field equations in light of the simplification (\ref{Lagrangian}). \ As
is well-known, there may be two distinct expressions for energy-momentum
tensors that result from any Lagrangian. \ From (\ref{Lagrangian}) the
canonical results for $\Theta _{\mu \nu }$, and its trace $\Theta =\Theta
_{\mu }^{\ \mu }$, are immediately seen to be%
\begin{equation}
\Theta _{\mu \nu }^{\left[ \text{canon}\right] }=\left( \partial _{\mu
}V_{\nu }\right) L^{\prime }-g_{\mu \nu }\left( au+L\right) \ ,\ \ \ \Theta
^{\left[ \text{canon}\right] }=vL^{\prime }-4\left( au+L\right) \ .
\label{TCanon}
\end{equation}%
Although not manifestly symmetric, it is nonetheless true that $\Theta _{\mu
\nu }^{\left[ \text{canon}\right] }=\Theta _{\nu \mu }^{\left[ \text{canon}%
\right] }$ \emph{on-shell} in light of the condition (\ref{GradientCond}). \ 

Surprisingly different results follow from covariantizing (\ref{Lagrangian})
with respect to an arbitrary background metric $g_{\mu \nu }$, varying the
action for $\sqrt{-\det g_{\alpha \beta }}~\mathcal{L}$ with respect to that
metric, and then taking the flat-space limit. \ This procedure gives the
\textquotedblleft gravitational\textquotedblright\ energy-momentum tensor
and its trace:%
\begin{equation}
\Theta _{\mu \nu }^{\left[ \text{grav}\right] }=-2\left( a+bL^{\prime
}\right) V_{\mu }V_{\nu }-g_{\mu \nu }\left( L-au-\left( v+2bu\right)
L^{\prime }\right) \ ,\ \ \ \Theta ^{\left[ \text{grav}\right] }=\left(
4v+6bu\right) L^{\prime }+2au-4L\ .  \label{TGrav}
\end{equation}%
The unusual structure exhibited in this tensor follows because in curved
spacetime $V^{\mu }$ as defined by (\ref{VandF}) is a \href{https://en.wikipedia.org/wiki/Tensor_density}%
{relative contravariant vector} of weight $+1$ with no dependence on the
metric, so $\partial _{\mu }V^{\mu }$ is a relative scalar of weight $+1$
also with no dependence on $g_{\mu \nu }$, and $V_{\mu }V^{\mu }=g_{\mu \nu
}V^{\mu }V^{\nu }$ is a relative scalar of weight $+2$ where all dependence
on the metric is shown explicitly. \ Hence the absolute scalar version of $%
\mathcal{L}\left( u,v\right) $ is given by 
\begin{equation}
\mathcal{L}=ag_{\mu \nu }V^{\mu }V^{\nu }/\left( -\det g_{\alpha \beta
}\right) +L\left( \left( \partial _{\mu }V^{\mu }\right) /\sqrt{-\det
g_{\alpha \beta }}+bg_{\mu \nu }V^{\mu }V^{\nu }/\left( -\det g_{\alpha
\beta }\right) \right) \ ,  \label{LinCurvedBackground}
\end{equation}%
where again all the metric dependence is shown explicitly.

It is straightforward to check on-shell conservation of either (\ref{TCanon}%
) or (\ref{TGrav}), separately. \ However, it turns out the flat-space
equations of motion can now be written in the form (\ref{Elegant}) provided
a linear combination of $\Theta _{\mu \nu }^{\left[ \text{canon}\right] }$
and $\Theta _{\mu \nu }^{\left[ \text{grav}\right] }$ is used for the
system's energy-momentum tensor. \ Let%
\begin{equation}
\Theta _{\mu \nu }=\frac{2}{3}~\Theta _{\mu \nu }^{\left[ \text{canon}\right]
}+\frac{1}{3}~\Theta _{\mu \nu }^{\left[ \text{grav}\right] }\ .
\label{NiceT}
\end{equation}%
The trace is then%
\begin{equation}
\Theta =\Theta _{\mu }^{\ \mu }=2\left( v+bu\right) L^{\prime }-4L-2au\ .
\label{NiceTrace}
\end{equation}

\subsection*{Field equation redux}

Since various scales have been previously chosen to set $m=1$, the field
equations (\ref{GradientCond}) and (\ref{SimpleFieldEqns}) give for the
left-hand side of (\ref{Elegant})%
\begin{equation}
\left( \square +1\right) V_{\mu }=\left( 1+\frac{1}{2}\frac{L^{\prime \prime
}}{a+bL^{\prime }}\right) \partial _{\mu }\left( v+bu\right) -b\partial
_{\mu }u\ ,
\end{equation}%
where (\ref{GradientCond}) implies $\square V_{\mu }=\partial ^{\lambda
}\partial _{\lambda }V_{\mu }=\partial ^{\lambda }\partial _{\mu }V_{\lambda
}=\partial _{\mu }v$. \ On the other hand, from (\ref{NiceTrace}) for any
constant $c$,%
\begin{equation}
c~\partial _{\mu }\Theta =2c\left( \left( v+bu\right) L^{\prime \prime
}-L^{\prime }\right) \partial _{\mu }\left( v+bu\right) -2ac\partial _{\mu
}u\ .
\end{equation}%
The choice $2ac=b$ reconciles the spurious $\partial _{\mu }u$ term to give
the desired form%
\begin{equation}
\left( \square +1\right) V_{\mu }=c~\partial _{\mu }\Theta
\label{ScaledFieldEqns}
\end{equation}%
\emph{provided} the function $L$ satisfies the second-order nonlinear
equation 
\begin{equation}
1+\frac{1}{2}\frac{L^{\prime \prime }\left( z\right) }{a+bL^{\prime }\left(
z\right) }=2c\left( zL^{\prime \prime }\left( z\right) -L^{\prime }\left(
z\right) \right) \ .  \label{ODE}
\end{equation}%
But note, the constant $c$ can be set to a convenient nonzero value by
further rescalings. \ \newpage

For example, if $\left( a,L\right) \ \rightarrow \ \left( \frac{ab}{2c},%
\frac{aL}{2bc}\right) $, along with the previous choice $2ac=b\ \rightarrow
\ a=1$, the equation for $L$ becomes%
\begin{equation}
1+\frac{1}{2b}\frac{L^{\prime \prime }}{b+L^{\prime }}=\frac{1}{b}\left(
zL^{\prime \prime }-L^{\prime }\right) \ .
\end{equation}%
Finally, rescaling $z\rightarrow w/b$ gives%
\begin{equation}
1+\frac{1}{2}\frac{L^{\prime \prime }}{1+L^{\prime }}=\left( wL^{\prime
\prime }-L^{\prime }\right)  \label{FinalODE}
\end{equation}%
The solution of this equation for $L^{\prime }$ with initial condition $%
L^{\prime }\left( 0\right) =0$ is%
\begin{equation}
L^{\prime }\left( w\right) =-1-2w+\sqrt{1+4w^{2}}\ .
\end{equation}%
Imposing the additional initial condition $L\left( 0\right) =0$, this
integrates immediately to%
\begin{equation}
L\left( w\right) =-w-w^{2}+\frac{1}{2}~w\sqrt{1+4w^{2}}+\frac{1}{4}~\ln
\left( 2w+\sqrt{1+4w^{2}}\right) \ .  \label{L(w)}
\end{equation}%
Comparison with (\ref{simple}) shows that%
\begin{equation}
L\left( w\right) =-w^{2}+2F\left( w\right) \ .
\end{equation}%
Given the previous rescalings, namely, $\mathcal{L}\left( u,v\right)
=au+L\left( v+bu\right) \rightarrow \left[ \frac{ab}{2c}u+\frac{1}{2bc}%
L\left( w=bz\right) \right] _{a=1}$, the Lagrangian density for the model
becomes%
\begin{eqnarray}
\mathcal{L}\left( u,v\right) &=&\frac{b}{2c}u+\frac{1}{2bc}\left( -bz-\left(
bz\right) ^{2}+\frac{1}{2}\left( bz\right) \sqrt{1+4\left( bz\right) ^{2}}+%
\frac{1}{4}~\ln \left( 2\left( bz\right) +\sqrt{1+4\left( bz\right) ^{2}}%
\right) \right)  \label{ExactL} \\
&=&\frac{b}{2c}u-\frac{b}{2c}z^{2}+\frac{b^{2}}{3c}z^{3}+O\left(
z^{4}\right) \ .
\end{eqnarray}%
As before, $v=\partial _{\mu }V^{\mu }$, $u=V_{\mu }V^{\mu }$, and $z=v+bu$.
\ Note that the term linear in $z$ in (\ref{ExactL})\ cancels out upon power
series expansion, so the result agrees with (\ref{IC}) up to and including
all terms of $O\left( V^{3}\right) $. \ 

To comport to the conventions in \cite{CF1980}, choose $b=-g$ and $c=g$, so
that $z=v-gu$, to find 
\begin{eqnarray}
\mathcal{L}\left( u,v\right) &=&-\frac{1}{2}~u-\frac{1}{2g^{2}}\left( 
\begin{array}{c}
g\left( v-gu\right) -g^{2}\left( v-gu\right) ^{2}-\frac{1}{2}g\left(
v-gu\right) \sqrt{1+4g^{2}\left( v-gu\right) ^{2}} \\ 
+\frac{1}{4}~\ln \left( -2g\left( v-gu\right) +\sqrt{1+4g^{2}\left(
v-gu\right) ^{2}}\right)%
\end{array}%
\right)  \label{FinalLagrangian} \\
&=&-\frac{1}{2}~u+\frac{1}{2}\left( v-gu\right) ^{2}+\frac{1}{3}~g\left(
v-gu\right) ^{3}+O\left( \left( v-gu\right) ^{4}\right) \ ,
\end{eqnarray}%
Now restore $m$ via the coordinate rescaling $x_{\mu }\rightarrow mx_{\mu }$%
, hence $v\rightarrow v/m$ and $\mathcal{L}\left( u,v\right) \rightarrow
m^{2}\mathcal{L}\left( u,v/m\right) $, thereby converting (\ref%
{ScaledFieldEqns}) into the form (\ref{Elegant}), with $\theta =m^{2}\Theta $%
.

\section*{Discussion}

The conventional integral equation form of (\ref{Elegant}), including a
free-field term with $\left( \square +m^{2}\right) V_{\mu }^{\left( 0\right)
}=0$, is given by%
\begin{equation}
V_{\mu }\left( x\right) =V_{\mu }^{\left( 0\right) }\left( x\right) +\frac{g%
}{m}\int G\left( x-y\right) ~\frac{\partial \Theta \left( y\right) }{%
\partial y^{\mu }}~d^{4}y\ ,  \label{ImplicitVSoln}
\end{equation}%
where $\Theta \left( y\right) $ depends implicitly on the field $V_{\nu
}\left( y\right) $ and $G$ is the usual isotropic, homogeneous, Dirichlet
boundary condition Green function that solves $\left( \square +m^{2}\right)
G\left( x-y\right) =\delta ^{4}\left( x-y\right) $. \ The free-field term
must be a gradient, $V_{\mu }^{\left( 0\right) }\left( x\right) =\partial
_{\mu }\Phi ^{\left( 0\right) }\left( x\right) $ with $\left( \square
+m^{2}\right) \Phi ^{\left( 0\right) }=0$, to ensure that $V_{\mu }\left(
x\right) =\partial _{\mu }\Phi \left( x\right) $ is also a gradient. \
Integration by parts followed by an overall integration then gives%
\begin{equation}
\Phi \left( x\right) =\Phi ^{\left( 0\right) }\left( x\right) +\frac{g}{m}%
\int G\left( x-y\right) ~\Theta \left( y\right) ~d^{4}y\ ,
\label{ImplicitPhiSoln}
\end{equation}%
where now $\Theta \left( y\right) $ depends implicitly on $\Phi \left(
y\right) $. \ That is to say, $\left( \square +m^{2}\right) \Phi =\frac{g}{m}%
\Theta \left[ \Phi \left( x\right) \right] $. \ 

On the one hand, this is not surprising, since there is a long-known
construction of an explicit local Lagrangian that leads directly to this
form for the scalar field equations \cite{FN}. \ (It amounts to the
Goldstone model after scalar field redefinition.) Taking a gradient to
reverse the steps above then leads back to (\ref{ImplicitVSoln}). \ On the
other hand, it is far from obvious that $\Theta \left[ \Phi \left( x\right) %
\right] $ can be re-expressed as a \emph{local} function of $V^{\mu
}=\partial _{\mu }\Phi $, and that $\Theta \left[ V^{\mu }\left( x\right) %
\right] $ follows in turn from a local, closed-form Lagrangian for $V^{\mu }$%
. \ The main point of this paper was to show that, indeed, there is an $%
\mathcal{L}$\ such that all this is true.

Were $\Theta $ due to anything other than $V^{\mu }$, field equations of the
form (\ref{Elegant}) would easily follow from%
\begin{equation}
\mathcal{L}_{\text{easy}}=\frac{1}{2}\left( \partial _{\mu }V_{\nu }\partial
^{\mu }V^{\nu }-m^{2}V_{\nu }V^{\nu }\right) +\frac{g}{m}V^{\mu }\partial
_{\mu }\Theta ^{\left[ \text{other}\right] }\ ,
\end{equation}%
i.e. a simple direct coupling of the vector to the gradient of any other
traced energy-momentum tensor. \ With a pinch of plausibility, this calls to
mind the axion coupling, albeit without the group theoretical and
topological underpinnings, not to mention \href{http://pdg.lbl.gov/2018/reviews/rpp2018-rev-axions.pdf}%
{the phenomenology}. \ 

In any case, Peter and I certainly did \emph{not} have axions in mind in
1980 when we wrote \cite{CF1980}. \ As best I can recall, we had only some
embryonic thoughts about massive gravity. \ In that context we speculated
(see \cite{CF1980}\ p 418) that $g/m\sim L_{\text{Hubble}}L_{\text{Planck}%
}=\left( 4.7\times 10^{-5}\unit{m}\right) ^{2}=1/\left( 4.2\times 10^{-3}%
\unit{eV}\right) ^{2}$. \ In retrospect, we were both struck by the fact
that this guess is approximately the same as phenomenological lower limits
for $1/m_{\text{axion}}^{2}$.

There is one more noteworthy piece of unfinished business in \cite{CF1980},
namely, a closed-form Lagrangian for a massive spin 2 field coupled to the
four-dimensional curl of its own energy-momentum tensor, where the spin 2
field is not the usual symmetric tensor, but rather the rank three tensor $%
T_{\left[ \lambda \mu \right] \nu }$ \cite{C}. \ For progress on this
additional unfinished business, please see \cite{CA}. \ With enough effort,
perhaps a complete formulation of this spin 2 model will also be available
soon, along with a few other variations on the theme of fields coupled to $%
\Theta _{\mu \nu }$. \ 

In closing, so far as I can tell, Peter had little if any interest in
totally antisymmetric tensor gauge fields prior to our paper \cite{CF1980}.
\ But he quickly pursued the subject in stellar fashion with his subsequent
work on dimensional compactification \cite{FR}. \ While all this work is
still conjectural, at the very least it provided and continues to provide
fundamental research problems in theoretical physics, especially for
doctoral students.

\subsection*{Acknowledgements}

I thank H. Alshal and D. Fairlie for comments and discussions, and the late
Peter Freund for \href{https://cgc.physics.miami.edu/PGOFMemories.pdf}{many
fond memories}. \ This work was supported in part by a University of Miami
Cooper Fellowship.


\begin{thebibliography}{9}
\bibitem{CF1980} T.L. Curtright and P.G.O. Freund, \textquotedblleft Massive
Dual Fields\textquotedblright\ \href{https://doi.org/10.1016/0550-3213(80)90174-1}%
{Nucl.Phys. B172 (1980) 413-424}.

\bibitem{FR} P.G.O. Freund and M. Rubin, \textquotedblleft Dynamics of
Dimensional Reduction\textquotedblright\ \href{https://doi.org/10.1016/0370-2693(80)90590-0}%
{Phys.Lett. 97B (1980) 233-235}.

\bibitem{WH} In other words, $W\left[ \mathcal{L}_{u},\mathcal{L}_{v}\right]
=0$, where $W\left[ f,g\right] =\det \left( 
\begin{array}{cc}
\partial f/\partial u & \partial g/\partial u \\ 
\partial f/\partial v & \partial g/\partial v%
\end{array}%
\right) $. \ Note that $W\left[ f,g\right] =0$ is a necessary (but \emph{not 
}sufficient) condition for a \emph{linear} dependence between $f$ and $g$ 
\cite{KW}. \ In fact, $W\left[ f,g\right] =0$ for \emph{any}
(differentiable) functional dependence between $f$ and $g$. \ For the case
at hand, $W$ is actually a determinant of\ a $2\times 2$ Hessian matrix, an
expression familiar from galileon models (e.g. see \cite{CF2012}).

\bibitem{KW} K. Wolsson, \textquotedblleft Linear dependence of a function
set of m variables with vanishing generalized Wronskians\textquotedblright\ 
\href{https://doi.org/10.1016/0024-3795(89)90548-X}{Linear Algebra and its
Applications 117 (1989) 73-80}.

\bibitem{CF2012} T.L. Curtright and D.B. Fairlie, \textquotedblleft A
Galileon Primer\textquotedblright\ \href{https://arxiv.org/abs/1212.6972}{%
arXiv:1212.6972 [hep-th]}.

\bibitem{FN} P.G.O. Freund and Y. Nambu \textquotedblleft Scalar Fields
Coupled to the Trace of the Energy-Momentum Tensor\textquotedblright\ \href{https://doi.org/10.1103/PhysRev.174.1741}%
{Phys.Rev. 174 (1968) 1741-1743}.

\bibitem{C} T.L. Curtright, \textquotedblleft Generalized Gauge
Fields\textquotedblright\ \href{https://doi.org/10.1016/0370-2693(85)91235-3}%
{Phys.Lett. 165B (1985) 304-308}.

\bibitem{CA} T.L. Curtright and H. Alshal, \textquotedblleft Massive Dual
Spin 2 Revisited\textquotedblright\ \href{https://arxiv.org/abs/1907.11532}{%
arXiv:1907.11532 [hep-th]}.
\end{thebibliography}
\end{document}